\documentclass[footinbib,aps,pra,reprint,superscriptaddress]{revtex4-1}
\usepackage{amsmath,amssymb,braket,upgreek,bm,verbatim}
\usepackage{graphicx,tabularx} 
\usepackage{float} %
\usepackage{xcolor}

\usepackage[absolute]{textpos}
\usepackage{xcolor}
\usepackage{multirow}
\usepackage[colorlinks=true,bookmarks=false,citecolor=blue,urlcolor=blue]{hyperref}
\usepackage{lineno}
\usepackage{blindtext}
\usepackage[absolute]{textpos}
\usepackage{physics}
\usepackage{amssymb}
\usepackage[T1]{fontenc}
\usepackage{orcidlink}

\begin{document}
\title{Resilient Entanglement Distribution in a Multihop Quantum Network}

\author{Muneer~Alshowkan}
\email{alshowkanm@ornl.gov}
\affiliation{Quantum Information Science Section, Oak Ridge National Laboratory, Oak Ridge, Tennessee 37831, USA}

\author{Joseph M. Lukens}
\affiliation{Quantum Information Science Section, Oak Ridge National Laboratory, Oak Ridge, Tennessee 37831, USA}
\affiliation{Research Technology Office and Quantum Collaborative, Arizona State University, Tempe, Arizona 85287, USA}

\author{Hsuan-Hao~Lu}
\affiliation{Quantum Information Science Section, Oak Ridge National Laboratory, Oak Ridge, Tennessee 37831, USA}

\author{Nicholas A. Peters}
\affiliation{Quantum Information Science Section, Oak Ridge National Laboratory, Oak Ridge, Tennessee 37831, USA}

\begin{abstract}
The evolution of quantum networking requires architectures capable of dynamically reconfigurable entanglement distribution to meet diverse user needs and ensure tolerance against transmission disruptions. We introduce multihop quantum networks to improve network reach and resilience by enabling quantum communications across intermediate nodes, thus broadening network connectivity and increasing scalability. We present multihop two-qubit polarization-entanglement distribution within a quantum network at the Oak Ridge National Laboratory campus. Our system uses wavelength-selective switches for adaptive bandwidth management on a software-defined quantum network that integrates a quantum data plane with classical data and control planes, creating a flexible, reconfigurable mesh. Our network distributes entanglement across six nodes within three subnetworks, each located in a separate building, optimizing quantum state fidelity and transmission rate through adaptive resource management. Additionally, we demonstrate the network's resilience by implementing a link recovery approach that monitors and reroutes quantum resources to maintain service continuity despite link failures---%
paving the way for scalable and reliable quantum networking infrastructures.
\end{abstract}

\maketitle
\begin{textblock}{13.3}(1.4,15.2)
\noindent\fontsize{7}{7}\selectfont \textcolor{black!30}{This manuscript has been co-authored by UT-Battelle, LLC, under contract DE-AC05-00OR22725 with the US Department of Energy (DOE). The US government retains and the publisher, by accepting the article for publication, acknowledges that the US government retains a nonexclusive, paid-up, irrevocable, worldwide license to publish or reproduce the published form of this manuscript, or allow others to do so, for US government purposes. DOE will provide public access to these results of federally sponsored research in accordance with the DOE Public Access Plan (http://energy.gov/downloads/doe-public-access-plan).}
\end{textblock}

\section{Introduction}
Quantum networks represent an advanced technological domain critical for utilizing and distributing quantum resources~\cite{Kimble2008}, %
enabling tasks beyond classical systems' capabilities~\cite{Bennett2014, Bennett1993, Giovannetti2004, Giovannetti2011}. A fundamental function of these networks is to establish entanglement between distant participants, essential for improved metrology~\cite{Giovannetti2004, Giovannetti2011, Bollinger1996}, distributed quantum computing~\cite{Cirac1999, Broadbent2009}, and improved security~\cite{Bennett1992, Gisin2007}. Future quantum networks must have the flexibility to distribute entanglement on-demand to various end-users with diverse resource needs, while navigating and mitigating unanticipated disruptions in transmission channels.

To this end, multihop quantum networks---analogous to their forerunners in the classical domain~\cite{Cerf1974, Kahn1978}---will be crucial for %
enabling extended network reach and improved robustness by facilitating %
versatile connectivity beyond direct point-to-point links. This advance will expand quantum networks' applicability and scalability and safeguard against service disruptions, enabling a more interconnected and adaptable infrastructure that will ultimately form a critical piece of a global quantum internet~\cite{Wehner2018,Pant2019}. %

Notable progress has been made in classically assisted multihop quantum key distribution (QKD), including trusted-relay QKD, where quantum data are converted to classical bits at trusted intermediary nodes~\cite{Elliott2002, Peev2009, Stucki2011, Evans2021}, as well as automatic link reestablishment after dynamic network reconfiguration~\cite{Chapuran_2009}. %
Additionally, the principles of software-defined networking (SDN) have been extended to quantum networks~\cite{Aguado2017,  Emilio2019, Aguado2019, Kozlowski2020, Alshowkan2022SMC, Chung2022}, yet the potential for SDN in multihop entanglement distribution has not been explored. In this work, we present a first demonstration to help close this critical gap as we expect larger scale networks will need to manage entanglement through many network hops.

Previous entanglement distribution experiments have focused on single-hop links with either \emph{fixed} or \emph{reconfigurable} connectivity. %
Fixed point-to-point entanglement distribution has utilized single and multichannel dense wavelength division multiplexers (DWDMs)~\cite{Autebert2016, Vergyris2019}, cascades thereof~\cite{wengerowsky2018, Joshi2020}, or even passive couplers%
~\cite{Townsend1997,Chapman2024}, while reconfigurable single-hop networks %
have used spatial~\cite{Herbauts2013, Laudenbach2020} 
or %
wavelength-selective switches (WSSs)~\cite{Lingaraju2021, Appas2021, Alshowkan2021, Che2022, Alshowkan2022b, Alshowkan2022c, Lu2024, miloshevsky2024} for more flexible and adaptive communication. 
Notwithstanding the \emph{spectral} reconfigurability facilitated by the latter case, all entanglement distribution experiments so far %
have been limited to one \emph{physical} lightpath between source and receiver. Thus, any given connection is vulnerable to the failure of a single fiber, requiring physical reconnection to resume service after a disruption.

In this paper, we implement multihop two-qubit entanglement distribution within a quantum network over the optical fiber infrastructure at Oak Ridge National Laboratory (ORNL). We use adaptive bandwidth management on a quantum-compatible SDN that integrates a quantum data plane with classical data and control planes, creating a flexible, reconfigurable mesh. Our network distributes entanglement across six nodes within three distant subnetworks, optimizing quantum state fidelity and transmission rate through adaptive resource management. Additionally, by incorporating redundant fiber lightpaths via transparent optical switches, we demonstrate the network's resilience to link failures, 
successfully monitoring and rerouting quantum signals through alternative paths
when the direct fiber route is blocked.
This work %
should pave the way for scalable and robust infrastructures combining SDN, flex-grid, and link redundancies for resource-efficient and resilient quantum communications. 

\begin{figure}[t!]
    \centering
    \includegraphics[width=0.35\textwidth]{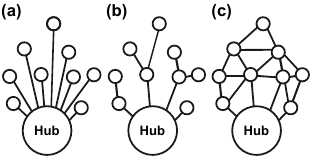}
    \caption{Classical network topologies. (a)~Single hop, where a single hub relays signals between nodes. (b)~Multihop, where intermediate nodes also act as hubs to relay traffic. (c) Multihop mesh, where additional node-to-node connections create redundant pathways between any given node and the hub.}
    \label{fig:placeholder1}
\end{figure}

\begin{figure*}[t!]
    \centering
    \includegraphics[width=\textwidth]{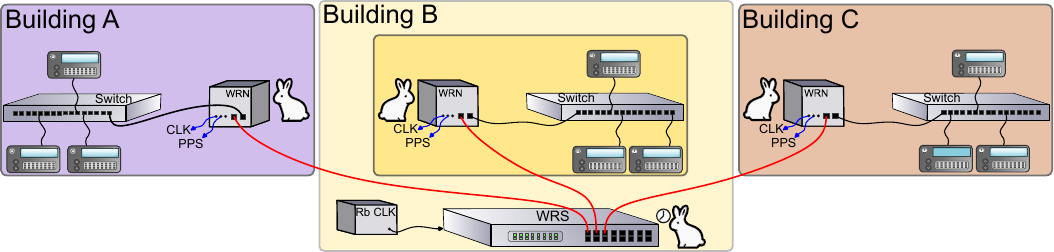}
    \caption{Clock distribution and classical communications across subnetworks using White Rabbit, with each subnetwork located in a seperate building. %
    CLK: output 10~MHz reference clock signal.
    PPS: output pulse per second.
    Rb CLK: rubidium frequency standard atomic clock.
    Switch: ethernet switch.
    WRN: White Rabbit node.
    WRS: White Rabbit switch.
    Black lines: ethernet signals (electrical).
    Blue lines: timing signals (electrical).
    Red lines: WR signals (optical).
    }
    \label{fig:Clocks}
\end{figure*}

\section{Multihop Networking}
\subsection{Multihop Classical Networking}
\label{sec:multihop}
Unlike single-hop transmission [Fig.~\ref{fig:placeholder1}(a)], where signals propagate directly from the source to destination, potentially mediated via a single hub, a multihop network [Fig.~\ref{fig:placeholder1}(b)] leverages intermediate nodes to reduce the number of hub connections without sacrificing communications potential. Moreover, by linking nodes in a mesh [Fig.~\ref{fig:placeholder1}(c)], such a multihop network can attain redundancy as well, wherein
intermediate nodes and devices such as routers and switches can select the most suitable paths based on %
traffic conditions, link availability, and other critical communications factors~\cite{Ramanathan2002}.

The multihop approach's practicality and utility are particularly evident when network resources are constrained or direct paths between communication endpoints are infeasible. For instance, in networks that span large geographical areas with scattered nodes and limited transmission ranges, establishing direct connections between all devices is often impractical or impossible. %

Moreover, %
by allowing data to hop from one node to another, networks can efficiently manage bandwidth, reduce congestion, %
and enhance overall communication reliability and performance. %
These networks can be either infrastructure-based, like a corporate wide-area network where data travels through multiple router hops, or infrastructure-less, such as a mobile or vehicular ad hoc network~\cite{Conti2007, Hartenstein2008}, or a combination of both as in a flying ad hoc network~\cite{Lakew2020}.

In ad hoc networks, traditional network roles %
are not applied in the conventional manner where routers forward data packets between different networks based on internet protocol (IP) addresses and switches route data within a single network using media access control (MAC) addresses. Instead, ad hoc networks leverage direct communications between nodes without the need for centralized infrastructure. In these networks, every node has the potential to perform \textit{routing} functions; it can receive, send, and forward data to other nodes based on the network's current topology and node locations. The traditional concept of a switch, which directs data within a network segment, is not applicable. Instead, nodes in ad hoc networks make routing decisions influenced by the network's protocol, considering factors like destination IP addresses, the most efficient path, and overall connectivity.
Thus, ad hoc networks %
are inherently more fluid and decentralized than traditional networks.

\subsection{Multihop quantum networking}
\label{sec:quantumMultihop}
In environments with fluctuating network conditions, the multihop approach provides the flexibility to adapt to changing topology and varying link quality. %
Such advantages are crucial for quantum networks, where the typically low light levels involved coupled with the unavailability of optical amplifiers %
make quantum signals arguably \emph{more} sensitive to link quality than their classical counterparts.
In infrastructure- and cost-limited scenarios, quantum networks can leverage multihop networking to achieve end-to-end communications through more accessible intermediate nodes. Indeed, this basic principle motivates much of satellite quantum communications, which has achieved recent successes in QKD~\cite{Pugh2017, Liao2018, Chen2021} and entanglement distribution~\cite{Liu2021, Liu2020}---links to intermediary satellites can bypass terrestrial impediments that would otherwise hamper transmission through a direct path.%

Moreover, link recovery will prove vital to the long-term reliability of future quantum networks. %
In case of link failure, the network must quickly adapt, rerouting quantum data through alternative paths to minimize %
 communications interruptions. This capability ensures a resilient network structure capable of handling uncertainties and maintaining operational efficiency. %
Beyond resiliency, %
multihop architectures enable connections between many nodes without requiring a fully connected mesh, which facilitates network expansion with minimal infrastructure overhaul.
Finally, %
multihop networks optimize valuable quantum resources by enabling shared connections and utilizing intermediate nodes to relay quantum information. %

In light of these considerations, we argue that multihop networking is indispensable for the development and practical deployment of quantum networks. By providing resiliency, supporting scalability, and ensuring efficient resource utilization, multihop techniques address many unique challenges faced by quantum communications.  %
Integrating and optimizing multihop capabilities within quantum networks, as well as within quantum computing platforms, will be crucial for realizing the vision of a global, interconnected quantum internet.

\section{Quantum Multihop Network Design}
\subsection{Timing Synchronization and Classical Communication}
\label{sec:timeSync}
White Rabbit timing synchronization, first applied for quantum networking in Ref.~\cite{Alshowkan2022b}, is employed in our testbed across three subnetworks in buildings A, B, and C shown in Fig.~\ref{fig:Clocks}.
This classical system %
not only distributes highly accurate clock signals but also supports ethernet communications. Consequently, we utilize the same infrastructure to manage the classical communications required for quantum networks. A rubidium (Rb) frequency standard (FS725; Stanford Research Systems)---a recent addition over the previous design in Ref.~\cite{Alshowkan2022b}---disciplines a White Rabbit switch (WRS; Safran) via a 10~MHz reference signal. The WRS then connects to a White Rabbit node (WRN; Safran) %
in each building via optical fiber connections using small form-factor pluggable (SFP) transceivers operating at 1310 and 1490~nm for bidirectional classical communications. %
Moreover, we use the additional SFP port on each WRN---originally intended for daisy chaining---to connect to a classical unmanaged ethernet switch. This configuration interconnects all node-supporting devices within the classical channel, making them accessible throughout the network.  
Finally, each WRN's 10~MHz output is doubled to 20~MHz using an arbitrary waveform generator (AWG; RIGOL). This signal, along with a pulse-per-second (PPS) output from the WRN, is fed into a time-to-digital converter (TDC) for precise global time stamping. The resulting synchronization infrastructure ensures accurate and reliable time distribution across the network nodes.

For data collection, we instruct all TDCs to record the time of detected events. Using a previously calibrated time delay---accounting for different path lengths photons travel to reach each detector---and a $\sim$1~ns coincidence window, we collect coincidences for each measurement setting between pairs of users. We perform these procedures for 36 polarization projections for quantum state tomography (QST) in rectilinear ($H/V$), diagonal ($D/A$), and circular ($R/L$) bases and subsequently feed this overcomplete dataset into a Bayesian QST workflow for quantum state estimation~\cite{Lukens2020, Lu2022b}.

\begin{figure*}[ht]
    \centering
    \includegraphics[width=\textwidth]{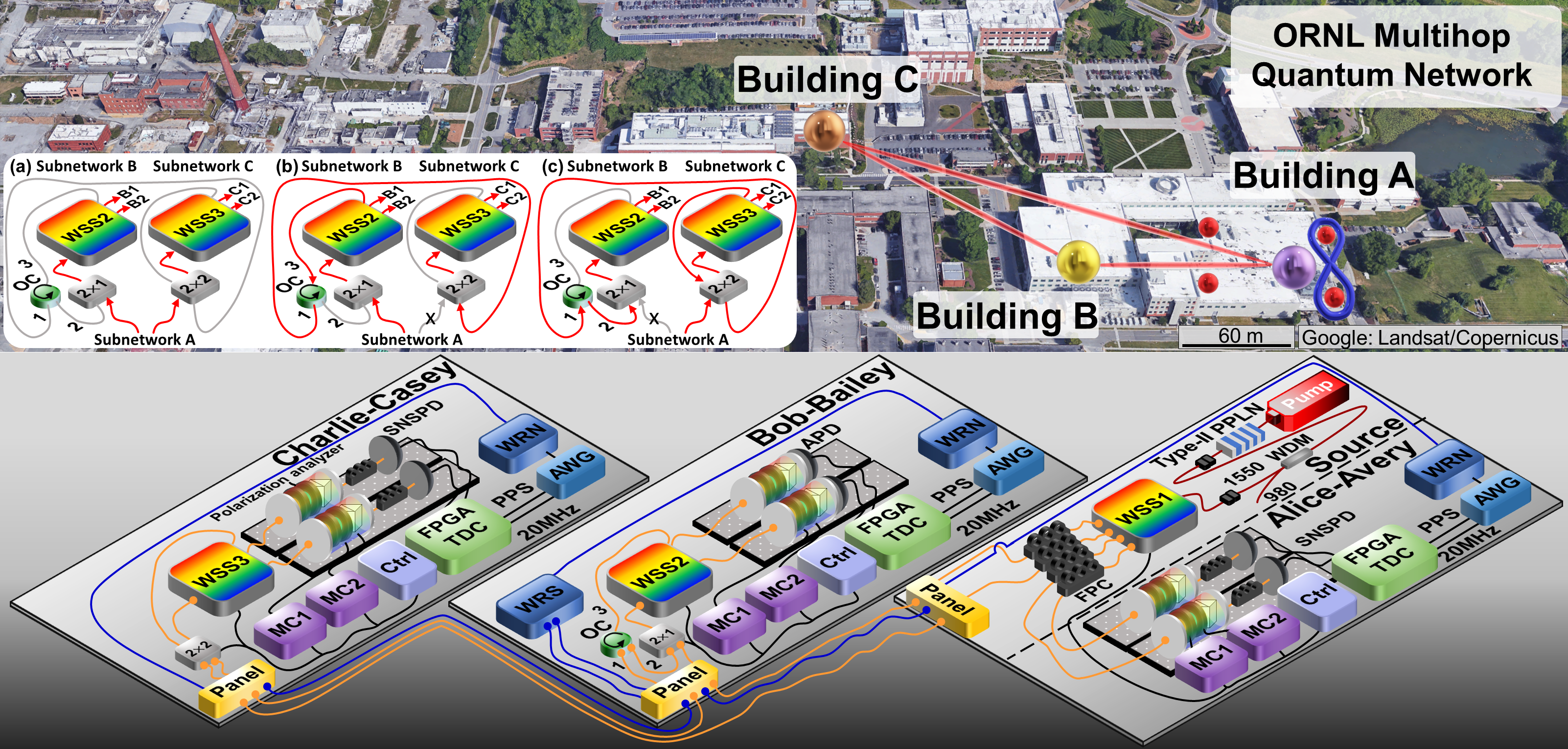}      
    \caption{Multihop network on ORNL campus. Bottom insets show the experimental setups in each building: blue (orange) lines represent the optical classical (quantum) signals, and black lines show the electrical timing and control signals. Top insets summarize the basic network configurations: red (gray) lines show active (inactive) signals for (a) direct connections from Alice to Bob and Charlie, (b) link recovery for building~C, and (c) link recovery for building~B.
    [APD: avalanche photodiode.
    AWG: arbitrary waveform generator.
    Ctrl: controller system.
    FPC: fiber polarization controller. 
    FPGA: field-programmable gate array.
    MC: motion controller. 
    OC: optical circulator.
    Panel: fiber-optic patch panel.
    PPLN: periodically poled lithium niobate. 
    PPS: pulse-per-second. 
    Pump: continuous-wave laser.
    SNSPD: superconducting nanowire single-photon detector. 
    TDC: time-to-digital converter.
    WDM: Wavelength division multiplexer.
    WRN: White Rabbit node.
    WRS: White Rabbit switch.
    WSS: wavelength-selective switch.
    2$\times$1: 2$\times$1 MEMS optical switch.
    2$\times$2: 2$\times$2 MEMS optical switch.]
    }
    \label{fig:multi-hop_exp}
\end{figure*}

\subsection{Software-Defined Quantum Networking}
Traditionally, tasks such as routing and data packet forwarding are integrated within the same device, such as a router or switch. However, the introduction of SDN separates the network's control plane~\cite{McKeown2008,Xia2015} from forwarding (or data plane), and thus decouples control decisions from data movement. %
The now-centralized control in the SDN plane can oversee and direct multiple forwarding devices, streamlining network operations and enabling quick adaptations to network congestion, device failures, or evolving network demand.
This architectural shift not only simplifies network management but also increases network programmability, making it easier to implement policy changes and integrate advanced technologies like cloud services and virtualization. %

The basic features of our quantum-compatible SDN architecture were introduced in Ref.~\cite{Alshowkan2022SMC}, which emphasizes a data plane comprising both classical and quantum devices. However, since all classical communications are now realized in a stand-alone fiber infrastructure isolated from the public network, the firewall and encryption devices mentioned therein are no longer required. Instead, the classical data plane now includes White Rabbit components and standard ethernet switches (Fig.~\ref{fig:Clocks}). The quantum data plane employs wavelength selective switches (WSSs; Finisar) of two types---single (1$\times$$9$) and dual (2$\times$1$\times$$20$)---both operating in the C-band, along with a 1$\times$$4$ pulse shaper (Waveshaper A4000; Finisar) that also functions as a WSS. These transparent optical devices facilitate the efficient sharing of optical resources through wavelength multiplexing and perform switching based on locally stored rules set by the controller. Such switching technologies based on liquid crystal on silicon 
for bandwidth allocation are supplemented with micro-electromechanical systems (MEMS) switches to form different physical network topologies. Both types of switching techniques involve no optical-to-electrical conversion. %

\section{Implementation}
\subsection{Deployed Network}
Figure~\ref{fig:multi-hop_exp} illustrates the multihop network, supporting the distribution of entangled photons across subnetworks in three buildings (A, B, and C) on the ORNL campus.
Building A holds the source and users Alice (A1) and Avery (A2), building B houses Bob (B1) and Bailey (B2), and building C is home to Charlie (C1) and Casey (C2). Each building has a fiber patch panel for interconnection with others. From each patch panel, fiber lightpaths pass through several telecommunication rooms before reaching their final destination, with total distances of 250~m (A$\rightarrow$B), 1.2~km (A$\rightarrow$C), and 930~m (B$\rightarrow$C).

A continuous-wave laser operated at 779.4~nm pumps approximately 25~mW into the pigtail of a periodically poled lithium niobate (PPLN) ridge waveguide (HC Photonics) designed for type-II spontaneous parametric downconversion (SPDC), ideally producing spectrally correlated, polarization-entangled photons in the Bell state $| \Psi^+\rangle\propto\ket{HV}+\ket{VH}$ (additional details can be found in Refs.~\cite{Lingaraju2021, Alshowkan2021}). The bandwidth generated by the source---with a full width at half-maximum of approximately 310~GHz---passes through a 980/1550~nm wavelength division multiplexer (WDM) to filter out the residual pump light and deliver the SPDC photons to WSS1, which %
divides the bandwidth into eight pairs of frequency-correlated channels (Chs. 1--8) with center frequencies $\omega_n = \omega_0 \pm \Delta\omega \left(n-\frac{1}{2}\right)$ for the signal (idler), where $\omega_0/2\pi = 192.3125$~THz and $\Delta\omega/2\pi = 25$~GHz---covering the 25~GHz-wide International Telecommunications Union channels from 21.25 through 25.00~\cite{ITU2020, Alshowkan2021}.

\begin{figure*}[t]
    \centering
    \includegraphics[width=\textwidth]{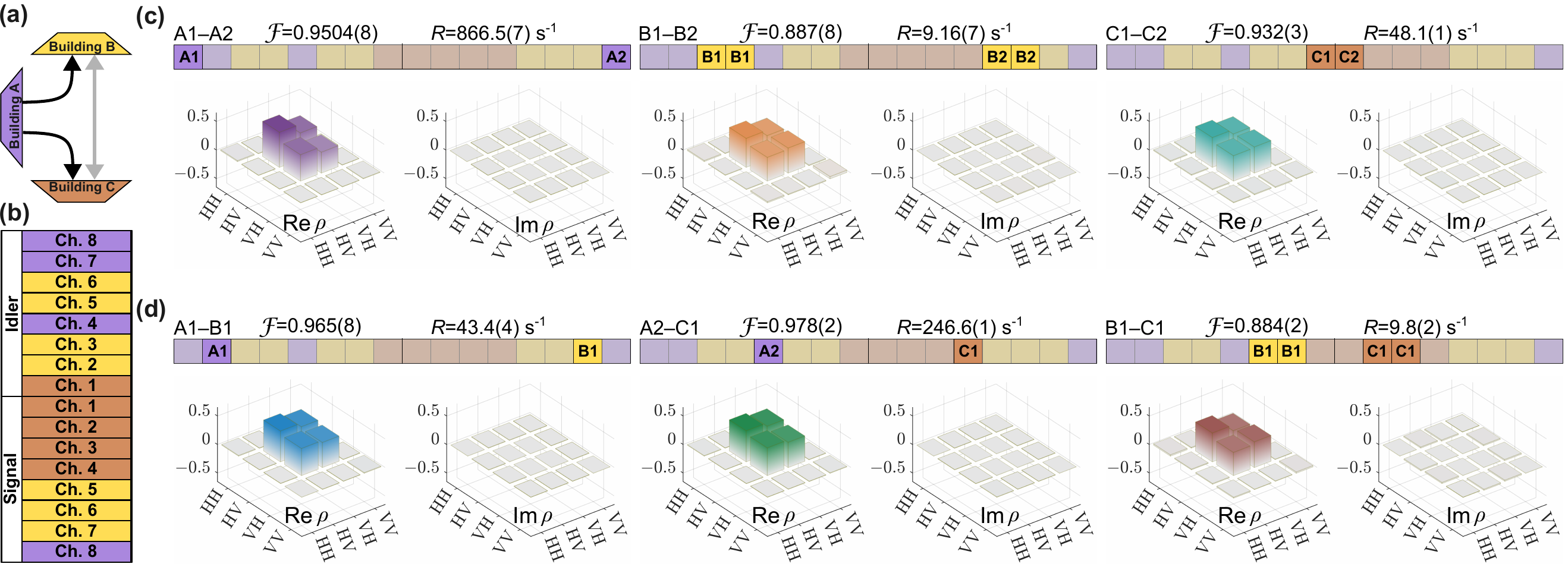}
    \caption{
    Direct links from building~A to B and C. (a) Optical lightpaths. (b) Spectral allocation. Colors indicate the terminating hubs for all frequency slots, whose center frequencies increase from 192.125~THz (idler Ch.~8) to 192.500~THz (signal Ch.~8) in 25~GHz steps. Density matrices, fidelities, and coincidence rates are estimated by Bayesian tomography for nodes connected to the same (c) and different (d) hubs. The frequency slots (left to right) in (c,d) correspond to the top-to-bottom ordering in (b), with user labels denoting the specific slots measured for the respective density matrix.
    }
    \label{fig:Direct_Links}
\end{figure*}

Two of WSS1's outputs are directed to users A1 and A2, while two other outputs are directed to a patch panel connected to buildings B and C.
The patch panel at B has two connections: one with building~A and another with building~C. The link from A is directed to a programmable 2$\times$1 MEMS switch, and that to C is connected to port~1 of a fiber-optic circulator: port 2 connects to the 2$\times$1 switch and port 3 to one of the WSS2 outputs.
Additionally, the output of the 2$\times$1 switch connects to the input of WSS2. %
Thus, the switch's passing state directs traffic from building~A to the input of WSS2, while the B$\rightarrow$C fiber path is connected to one of the WSS2 outputs.  In the crossing state, the fiber path C$\rightarrow$B becomes the input to WSS2, leaving the lightpath A$\rightarrow$B blocked. %

Similarly, the patch panel at C has two quantum connections, one to building~A and another to building~B. Both fibers are directed to a 2$\times$2 MEMS switch, whose outputs connect to the input and one of the outputs of WSS3. %
In the passing (crossing) state of the 2$\times$2 switch, the traffic from building~A (B) is directed to the input of WSS3, and the fiber linking building~B (A) is connected to one of the WSS3 outputs. 

The insets in Fig.~\ref{fig:multi-hop_exp}(a--c) depict the network configurations enabled by various states of the MEMS switches. In (a), photons from building~A reach B and C directly; in (b), the loss of the A$\rightarrow$C link is addressed by routing C's spectrum through a redundant multihop lightpath A$\rightarrow$B$\rightarrow$C; in (c), loss of the A$\rightarrow$B link is addressed by activating the A$\rightarrow$C$\rightarrow$B multihop path. This ensures the network's resilience against the loss of any single building-to-building connection, i.e., protection switching. %

In each location, the two users are connected to their respective WSS, which functions as a local hub. Each user has a polarization analyzer comprising a quarter-wave plate, half-wave plate, and polarizing beamsplitter, with each wave plate connected to a motorized motion controller. The outputs from the polarization analyzers are fed into single-photon detectors: superconducting nanowire single-photon detectors (SNSPDs; Quantum Opus) for A1, A2, C1, and C2 (with efficiency $>$81\%), and avalanche photodiodes (APDs; ID Quantique) for B1 and B2 (operated in free-running mode with efficiency set at 20\% and dead time set at 10~$\upmu$s). %
The detector output pulses are time-stamped by the FPGA-based TDCs, which are synchronized using the White Rabbit timing system. %

\subsection{Direct Links}
Our reconfigurable network enables various topologies. We first consider the simple star-type topology between the hubs as a benchmark, %
with WSS1 as the centralized hub and direct optical paths to WSS2 and WSS3, as shown physically in Fig.~\ref{fig:multi-hop_exp}(a) and schematically in Fig.~\ref{fig:Direct_Links}(a). This creates a simple network that follows the hub-and-spoke communication paradigm~\cite{Metcalfe1976}, resulting in the lowest loss between the users. %

Two general classes of connections can be formed to distribute entanglement, distinguished by whether the entangled users share the same WSS hub (intrabuilding) or different WSS hubs (interbuilding). %
Figure~\ref{fig:Direct_Links}(b) highlights one possible bandwidth allocation supporting all such connections, where the color of each 25~GHz slot corresponds to the destination building. %
The SDN application layer %
sends commands to each WSS, %
instructing WSS1 to keep its allocated bandwidth and send to WSS2 and WSS3 their allocations. %
Then, each WSS distributes slices to the respective users. Additionally, our SDN controller sends commands to the MEMS switches programming them to the passing state to direct the traffic originating from WSS1 to the inputs of WSS2 and WSS3. Characterizing our links reveals that the total loss from the input of WSS1 to the output of the polarization analyzer of each user is as follows: 7.25~dB (A1), 8.51~dB (A2), 17.4~dB (B1), 16.6~dB (B2), 17.2~dB (C1), and 17.6~dB (C2).

For the intrabuilding links, average singles count rates are (in units of s$^{-1}$): 44,497(3) and 35,164(2) (A1--A2); 11,415(1) and 14,201(1) (B1--B2); 18,170(1) and 18,092(1) (C1--C2). %
Performing QST concurrently between all users with an integration time of 180~s per setting, Bayesian estimation yields %
the results in 
Fig.~\ref{fig:Direct_Links}(c), with the following fidelities [0.9504(8), 0.887(8), 0.932(3)] with respect to $|\Psi^{+}\rangle$ and coincidence rates [866.5(7), 9.16(7), 48.1(1)] s$^{-1}$ for these three links. %
For the interbuilding links, %
we obtain
average singles count rates (in units of s$^{-1}$): 49,409(7) and 5,935(2) (A1--B1); 102,010(10) and 10,366(3) (A2--C1); 13,253(4) and 10,617(3) (B1--C1). %
Bayesian QST with a 30~s integration time per setting produces 
the results in Fig.~\ref{fig:Direct_Links}(d) with fidelities [0.965(8), 0.978(2), 0.884(2)] and coincidence rates [43.4(4), 246.6(1), 9.8(2)] s$^{-1}$ for these three links. %

\subsection{Link Recovery via Building B}

\begin{figure}[t!]
    \centering
    \includegraphics[width=0.4\textwidth]{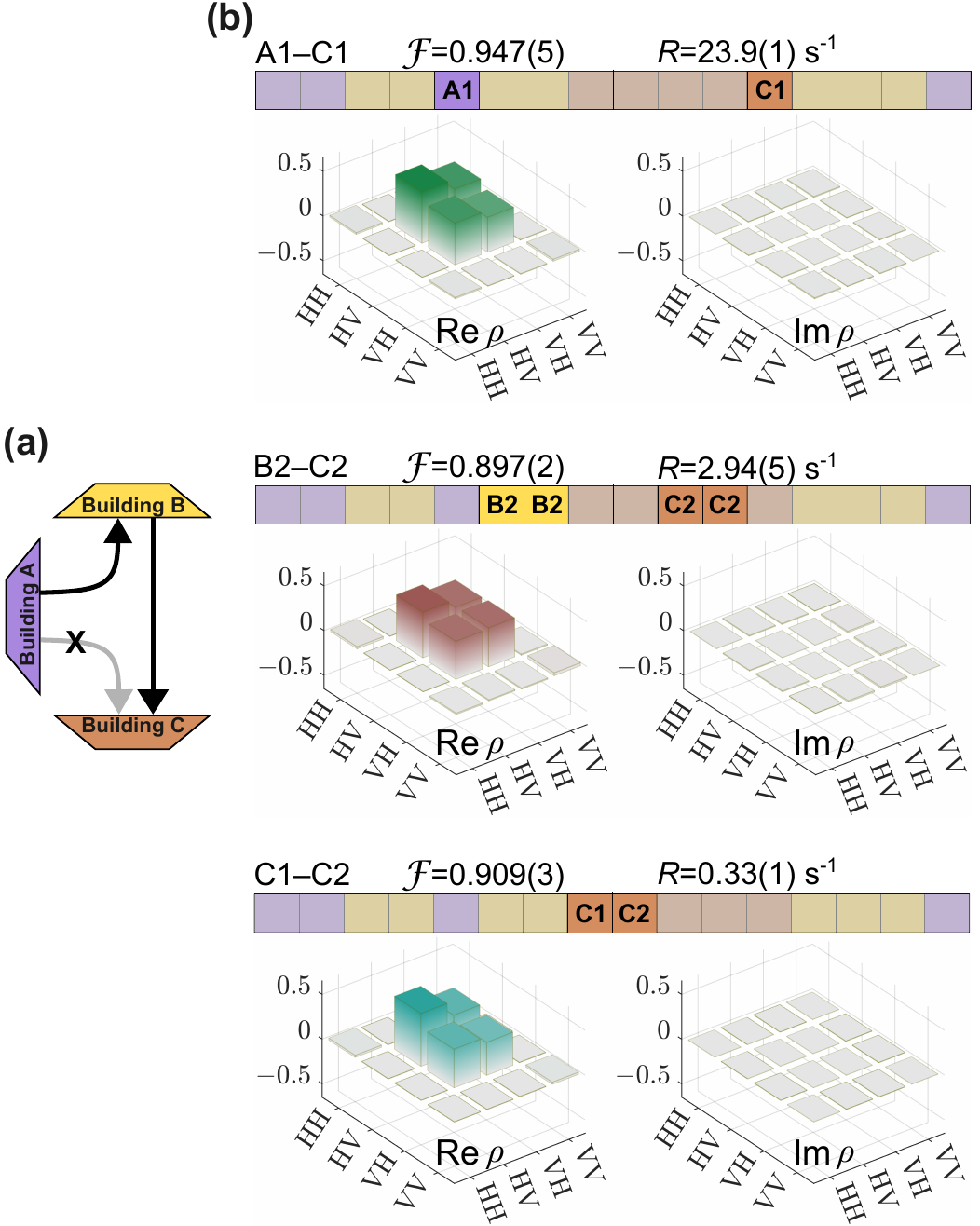}    
    \caption{
    Link recovery via building~B when the A$\rightarrow$C connection has been broken. (a) Optical lightpaths. Spectrum addressed to C is now rerouted via B. %
    (b) Experimental results. The spectral allocation is identical to Fig.~\ref{fig:Direct_Links}(b).
    }
    \label{fig:Recovery_via_B}
\end{figure}

In this scenario, we consider building A losing its direct fiber connection to building C (A$\rightarrow$C is out of service). As a mitigation strategy, communication is routed to C via building B, creating a more complex topology than a typical star network [Figs.~\ref{fig:multi-hop_exp}(b) and \ref{fig:Recovery_via_B}(a)]. Quantum signals now travel to WSS3 via WSS2 (i.e., A$\rightarrow$B$\rightarrow$C). %
While this configuration ensures continuous operation, it also introduces losses, with the total loss from the WSS1 input to the polarization analyzer output of C1 (C2) now measuring 27.5~dB (26.8~dB) due to this additional hop---approximately 10 dB more than the original direct A$\rightarrow$C path.

To enable this configuration, the SDN application layer sets a threshold (determined by dark counts and background noise) for the minimum output singles counts for each detector. %
When the singles counts drop below the threshold, %
the controller directs WSS1 to send all building~C bandwidth slots to building B and WSS2 to utilize its designated bandwidth and forward the remaining bandwidth via an output fiber to WSS3. 
At building~C, the 2$\times$2 MEMS switch
changes to the crossing state, %
making traffic from building B the input to WSS3. %
This reestablishes the disrupted entanglement links between A1--C1, B2--C2, and C1--C2. %

Following the same bandwidth allocation plan as in Fig.~\ref{fig:Direct_Links}(b), but now with all C slots passing through B first, %
average singles count rates (s$^{-1}$) are as follows: 88,759(5) and 1,281(1) (A1--C1); 17,041(2) and 2,650(1) (B2--C2); 1,661(1) and 1,613(1) (C1--C2). %
Bayesian QST with an integration time per measurement of 120~s for A1--C1 and B2--C2 and 180~s for C1--C2 returns the results in Fig.~\ref{fig:Recovery_via_B}(c): fidelities [0.947(5), 0.897(2), 0.909(3)] and coincidence rates [23.9(1), 2.94(5), 0.33(1)] s$^{-1}$ for these three links. %
Comparing the recovered links in Fig.~\ref{fig:Recovery_via_B}(c) to their building equivalents in Fig.~\ref{fig:Direct_Links}(c,d)---i.e., A1--C1 here with A2--C1 there, B2--C2 here with B1--C1 there, and C1--C2 in both---confirms good performance. Fidelities are comparable to within a few percentage points, and the reductions in coincidence rates are in generally good agreement with expectations for the increased multihop loss, with the C1--C2 link suffering the greatest hit as both photons traverse the multihop path.

\begin{figure}[t!]
    \centering
    \includegraphics[width=0.4\textwidth]{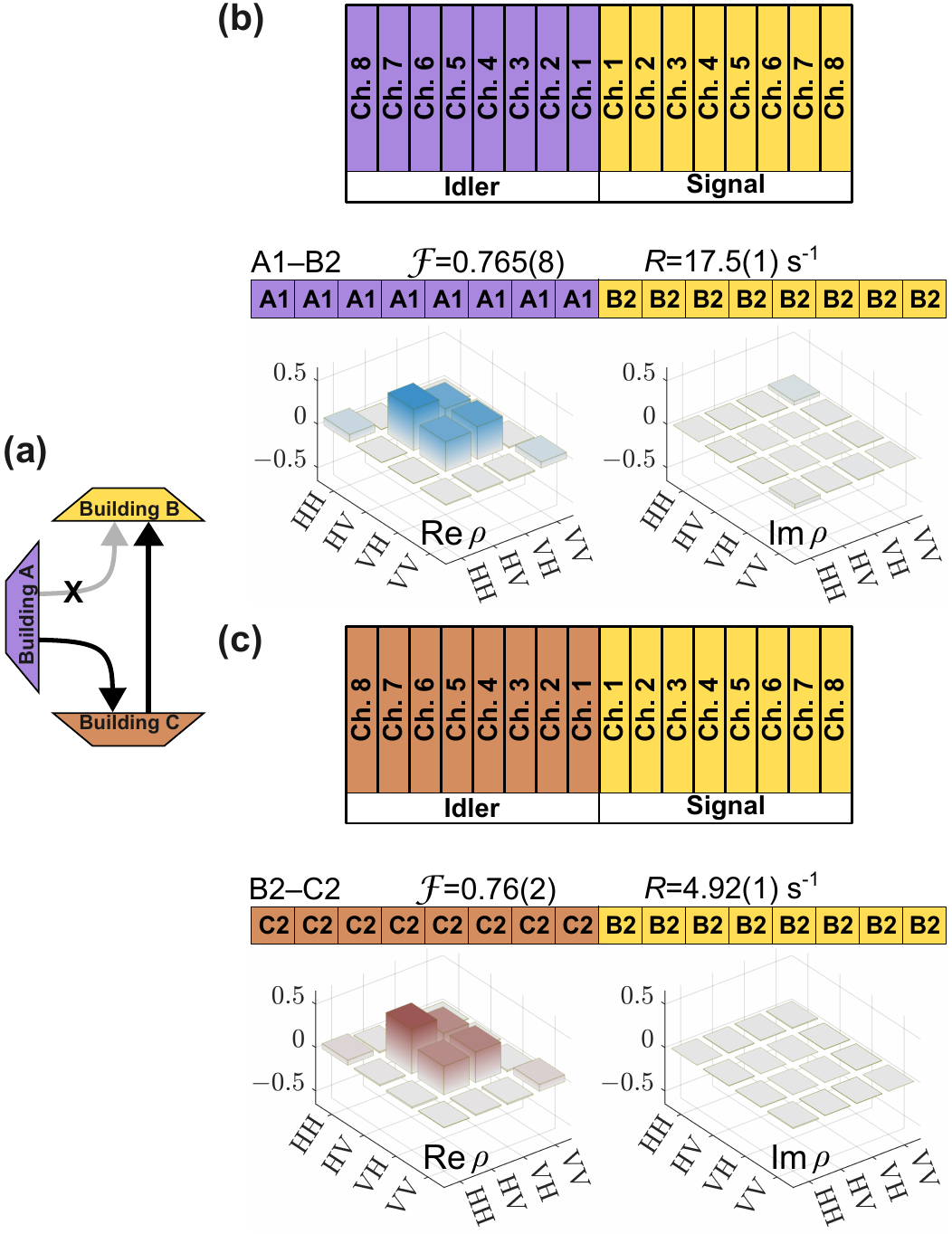}    
    \caption{
    Link recovery via building~C when the A$\rightarrow$B connection has been broken. (a) Optical lightpaths. Spectrum addressed to B is now rerouted via C. Experimental results for the interbuilding links (b) A1--B2 and (c) B2--C2. As before, the channel frequencies from left to right increase from 192.125~THz (idler Ch.~8) to 192.500~THz (signal Ch.~8) in 25~GHz steps.
    }
    \label{fig:Recovery_via_C}
\end{figure}

\subsection{Link Recovery via Building C}
We address another complex network topology where a direct optical fiber connection between buildings A and B has been lost. %
Here quantum traffic from WSS1 to building B is directed through WSS3 to reach WSS2 [Figs.~\ref{fig:multi-hop_exp}(c) and \ref{fig:Recovery_via_C}(a)]. To form this configuration, the SDN controller issues commands to the 2$\times$1 (2$\times$2) switch to enter the crossing (passing) state, activating a lightpath from A$\rightarrow$C$\rightarrow$B via WSS3.
In this recovered path, the total loss from the WSS1 input to B1 (B2) is now 28.8~dB (28.0~dB), again $\sim$10~dB higher than the original direct A$\rightarrow$B path.

Due to low detection efficiencies for %
APD-limited users in building~B, in lieu of the bandwidth allocation in Fig.~\ref{fig:Direct_Links}(b) we consider two mutually exclusive ``priority'' allocations where all available bandwidth is given to the two users of interest.
We first assign all eight channels to 
A1--B2 [Fig.~\ref{fig:Recovery_via_C}(b)], and then distribute all available bandwidth slices for communication between B2--C2 [Fig.~\ref{fig:Recovery_via_C}(c)]. In these two configurations, we find average singles count rates (s$^{-1}$) of 341,315(9) and 4,198(1) (A1--B2) and 96,091(5) and 4,071(1) (B2--C2). %
Bayesian QST with a 120~s integration time per measurement returns fidelities [0.765(8), 0.76(2)] and coincidence rates [17.5(1), 4.92(1)] s$^{-1}$ for links A1--B2 and B2--C2, respectively.

Due to the full-band allocations, the coincidence rates are able to compensate for much of the added loss, matching the direct link counterparts---A1--B1 and B1--C1 in Fig.~\ref{fig:Direct_Links}(d)---to within a factor of $\sim$2--3. %
However, the larger bandwidth increases the multipair emission noise during a given coincidence window, leading to lower fidelities of $\sim$0.76. Nonetheless, the states are still entangled, with log-negativity values of 0.62(2) ebits for A1--B2 and 0.61(3) ebits for B2--C2, which as upper bounds on distillable entanglement indicate their potential use in quantum applications~\cite{Vidal2002}.

These effects highlight the tradeoffs involved in extending network reach and enhancing redundancy through complex network topologies. While such configurations are beneficial for overcoming physical and infrastructure limitations, they require careful consideration of the impacts on quantum communications performance and fidelity. Our experiments show both the limitations and practical value of multihop quantum networking. For example, although the recovered links in Fig.~\ref{fig:Recovery_via_C} have significantly lower fidelity than their direct counterparts in Fig.~\ref{fig:Direct_Links}---and utilize all available bandwidth while doing so---they are \emph{possible}. Without multihop networking and a redundant fiber channel, building B would have no opportunity whatsoever to receive entanglement after loss of the A$\rightarrow$B fiber.
Overall, our results
emphasize the need for optimized multihop network design and state-of-the-art quantum systems to preserve entanglement quality across extended distances.

\section{Conclusion}
\label{sec:conclusion}
Leveraging adaptive bandwidth management on an SDN, we realize a reconfigurable multihop quantum network that integrates quantum and classical data planes and efficiently distributes entanglement to six users across three buildings. Furthermore, our approach to network resilience, utilizing redundant fibers and MEMS switches for link recovery, ensures continuous operation despite link failures. These advancements underscore the potential for scalable and robust networking infrastructures critical to future quantum networks. %

\section*{Acknowledgments}
We thank B.~J. Lawrie for sharing lab space and SNSPDs. We also thank C.~E. Marvinney, I. Gallagher, and J.~C. Chapman for helping with SNSPDs. This work was performed at Oak Ridge National Laboratory, operated by UT-Battelle for the U.S. Department of energy under contract no. DE-AC05-00OR22725.

\end{document}